\documentstyle[12pt,fullpage]{article}

\author{A.N.~Aleshaev, I.V.~Pinayev,  V.M.~Popik,  S.S.~Serednyakov,
\\T.V.~Shaftan, A.S.~Sokolov, N.A.~Vinokurov, P.V.~Vorob'ev}

\title{A Study of Influence of Synchrotron
               Radiation Quantum Fluctuations on the
               Synchrotron Oscillations of a Single
               Electron Using Undulator Radiation}
\date{ Budker Institute of Nuclear Physics, 11 Lavrentyev Ave., Novosibirsk,
 630090, Russian Federation}

\begin{document}

\maketitle

\abstract{A single electron circulating in a storage ring
is a very peculiar object. Synchrotron radiation quantum fluctuation
causes stohastic process in the synchrotron oscillation of an electron.
The radiation from a undulator permits one to obtain discret moments of
longitudional electron motion. Experiments with a single electron on
on the VEPP-3 optical klystrons are described.}

\section{Introduction}

     In the previous paper [1] we reported the results of the
Brown-Twiss experiment with undulator radiation of a single electron
circulated in a storage ring. It was found that the correlation length
of photocounts is  sufficiently  less  (actually,  less  than  the  time
resolution of our equipment) than the  ``natural''  bunchlength
excited by the quantum  fluctuations  of  synchrotron  radiation.  The
interpretation of this result is the following: the length of
localization of the electron  is  sufficiently  less  than  the  natural
bunchlength. It was also indicated that the longitudinal  coordinate  of
the electron oscillates with the synchrotron frequency.

     In this paper we describe the investigation of  the  single
electron longitudinal motion.

\section{The experimental set-up}

     The layout of our installation is shown in the Fig.~1,~2. The electron
circulates in the storage ring VEPP-3 with the revolution frequency
$F_{rev}=4.012$ MHz modulated by the frequency of synchrotron oscillations
$\Omega/2\pi$ (about 1 kHz). The light emitted by the electron in the
undulator
is detected by the photomultiplier. The pulses of photocounts pass through
the discriminator-shaper. These pulses of a standard shape give start to the
time-to-digital converter. The reference pulses phased with storage ring
RF system coming with revolution frequency give stop to the converter.
So, we measure the delay between  the  time  when the electron  enters to the
undulator and the first reference pulse after this  event.  The  average
frequency of the photocounts (about 20 kHz) is less than the  revolution
frequency, therefore another circuit  counts  the  number  of  revolution
periods between the photocounts. The delay and the number of revolutions
are writing into the memory of the intellectual CAMAC  controller.  Thus
the result of measurements is the couple of arrays which  comprises  the
dependence of delay on the revolution number. The  small  part  of  this
dependence is shown in Fig.~3.

\section{Preliminary data handling}

     To obtain more precize information about the electron dynamic,
we have to exclude the errors from the experimental data. There are two
sources of errors. First, uncorrelated photocounts, which are presents even
in the absence of an electron. Second, the jitter of pulses at the output
of discriminator-shaper caused by the statistical fluctuation of the  shape
of pulses from the photomultiplier. On the another hand, we know, that
the dependence of the longitudinal coordinate $s$ of an electron on time $t$
is sinusoidal with slowly varying amplitude $A$ and phase $\phi$:
\begin{equation}
               s = c t + A(t) sin(\Omega n/F_{rev}+\phi(t))
\end{equation}
where n is the revolution number, and $c$ is speed of light.

     Taking into  account  these  circumstances  we  use  the  following
algorithm of the filtering of the sequence of delays. We take the  small
part of sequence with the time duration $T$ (typically we choose $T$ about a
few
periods of the synchrotron oscillations). Using the least-squares
method we obtain the amplitude and the phase of the fitting sinusoid
and the rms error. After that we exclude from consideration all points
with the deviation more than two rms errors. We repeat such least square
fitting twice more, excluding about 15\% of experimental points and obtain
``clean'' sequence (Fig.~4) with fitted amplitude and phase of
oscillations.

     To estimate the ``optimal'' value of the partial sequence duration $T$
we used the consideration of the optimal filter for a sequence [2].
The least-squares procedure is actually the sort of filters and $T^{-1}$
is the width of the filter. One can easily obtain the estimation for the
optimal value of $T$
\begin{equation}
          T_{opt} \approx \frac{\sigma}{A}\sqrt{\frac{\tau}{\nu}}
\end{equation}
where $\sigma$ is the rms error, $\tau$ is the longitudinal dumping  time,
$\nu$ is the frequency of photocounts. Actually we chose $T$ few times more
than for average amplitude. So, we lost in the resolution for large
amplitudes, but decreased the noise dramatically for all amplitudes.

     Applying such treatment to various partial sequences we obtain
the time dependence of amplitude $A(t)$ and slow phase $\phi (t)$ on
time (Fig.~5). The electron trajectory in the polar coordinates
$A$, $\phi$ is shown in Fig.~6.

     Let us discuss qualitatively some features of this  trajectory.
It looks continuous but non-differentiable. These  properties  are  very
natural for the Brownian motion. This view of trajectory indicates also,
that our filtering  algorithm  provide  a  good  suppression  of  noises
(otherwise there would appear discontinuities)  and  rather  broad  band
(otherwise the trajectory would be smooth). The motion is irregular
except of large amplitudes where there is the regular rotation caused by
the nonlinearity of synchrotron oscillations. The trajectory fills
more or less homogeneously the central part of the phase plane except of
the close vicinity of the origin point. This ``small hole'' in the phase
space distribution may be caused by the fluctuations of the RF voltage
(see below).

\section{The correlation functions}

 The correlation functions for amplitude and phase are shown in Fig.~7,
and Fig.~8, correspondingly. The characteristic durations are very
different ($\approx$300~ms and $\approx$15~ms ). Note that the correlation
time for the amplitudes is almost equal to the damping time of
the synchrotron oscillations. The short correlation time for phase may be
also caused by the fluctuations of the RF voltage.

\section{Observations of two electrons.}

     To separate the influence of the quantum fluctuations of radiation
on an electron dynamics from the influence of another noises we
performed the same experiment with two electrons in the storage ring.
The ``kicks'' of the photon emission for two electrons are obviously
uncorrelated, but ``kicks'' from noise of the RF and magnetic systems are
correlated. The handling of the photocounts data is more complicated
                now and demands to perform the further investigations.

\section{Conclusion}

     In the experiments described in  this  paper  we  demonstrated the
stochastic behavior of  the  electron similar to  the  Brownian
motion. But the phenomenon is sufficiently different from the last one as
the temperature of the field oscillators which interacts  with  electron
is equal to zero. In the reference system moving with electron the effective
temperature  proportional  to  its  acceleration  appears [3]
and  the
interpretation becomes  similar  to  the  conventional  consideration  of
thermal fluctuations.

\newpage
\appendix
\pagestyle{empty}
{\large\bf Figures Caption}

\newcounter{N}
\begin{list}{Fig. \arabic{N}}{\usecounter{N}}
\item Layout of the installation

\item Layout of the time interval measurement

\item The small part of the measured dependence

\item The procedure of the data cleaning

\item Amplitude and phase of synchrotron oscillations

\item Trajectory of single electron in phase coordinates

\item The amplitude correlation

\item The phase correlation

\end{list}

\newpage

\begin{figure}
\unitlength=1.00mm
\special{em:linewidth 0.4pt}
\linethickness{0.4pt}
\begin{picture}(140.00,95.75)
\put(60.00,70.00){\oval(34.00,40.00)[l]}
\put(125.00,70.00){\oval(30.00,40.00)[r]}
\put(60.00,90.00){\vector(-4,1){23.00}}
\put(60.00,90.00){\vector(-4,-1){23.00}}
\put(60.00,90.00){\vector(-1,0){23.00}}
\put(19.00,86.00){\framebox(15.00,7.00)[cc]{PMT}}
\put(79.00,68.00){\makebox(0,0)[cc]{VEPP - 3}}
\put(60.00,86.00){\framebox(65.00,7.00)[cc]{UNDULATOR}}
\put(121.00,51.00){\line(4,-1){4.03}}
\put(125.03,49.92){\line(-4,-1){3.92}}
\put(85.00,40.00){\framebox(16.00,20.00)[cc]{  }}
\put(60.00,50.00){\line(1,0){25.00}}
\put(101.00,50.00){\line(1,0){24.00}}
\put(106.00,37.00){\makebox(0,0)[cc]{RF CAVITY}}
\put(88.00,40.00){\line(0,-1){3.87}}
\put(88.00,36.13){\line(-1,0){84.00}}
\put(21.00,86.00){\line(0,-1){34.82}}
\put(21.00,51.18){\line(-1,0){17.00}}
\put(5.00,54.00){\makebox(0,0)[cc]{START}}
\put(5.00,39.00){\makebox(0,0)[cc]{STOP}}
\end{picture}

\end{figure}

\newpage

\begin{figure}
\unitlength=1mm
\special{em:linewidth 0.4pt}
\linethickness{0.4pt}
\begin{picture}(145.00,100.00)
\put(5.00,90.00){\vector(1,0){15.00}}
\put(20.00,80.00){\framebox(10.00,20.00)[cc]{D}}
\put(5.00,50.00){\vector(1,0){15.00}}
\put(20.00,40.00){\framebox(10.00,20.00)[cc]{D}}
\put(45.00,75.00){\framebox(10.00,20.00)[cc]{\&}}
\put(30.00,90.00){\vector(1,0){15.00}}
\put(45.00,45.00){\framebox(25.00,11.00)[cc]{Delay}}
\put(30.00,50.00){\vector(1,0){15.00}}
\put(35.00,27.00){\line(0,1){53.00}}
\put(35.00,80.00){\vector(1,0){10.00}}
\put(100.00,45.00){\framebox(45.00,42.00)[cc]{}}
\put(55.00,80.00){\vector(1,0){45.00}}
\put(70.00,50.00){\vector(1,0){30.00}}
\put(100.00,20.00){\framebox(45.00,15.00)[cc]{Counter}}
\put(35.00,27.00){\vector(1,0){65.00}}
\put(96.00,82.00){\makebox(0,0)[rb]{start}}
\put(96.00,53.00){\makebox(0,0)[rb]{stop}}
\put(122.00,75.00){\makebox(0,0)[cc]{Time-to-Digit}}
\put(122.00,55.00){\makebox(0,0)[cc]{Convertor}}
\put(5.00,92.00){\makebox(0,0)[lb]{PMT}}
\put(5.00,52.00){\makebox(0,0)[lb]{4 MHz}}
\end{picture}
\end{figure}

\newpage

\begin{figure}
\unitlength=0.24pt

\special{em:graph i6.pcx}
\vspace{20cm}
\end{figure}

\newpage

\begin{figure}
\unitlength=0.24pt

\special{em:graph i5.pcx}
\vspace{20cm}
\end{figure}

\newpage

\begin{figure}
\unitlength=0.24pt

\special{em:graph i3.pcx}
\vspace{20cm}
\end{figure}

\newpage

\begin{figure}
\unitlength=0.24pt

\special{em:graph i4.pcx}
\vspace{20cm}
\end{figure}

\newpage

\begin{figure}
\unitlength=0.24pt

\special{em:graph i2.pcx}
\vspace{20cm}
\end{figure}

\newpage

\begin{figure}
\unitlength=0.24pt

\special{em:graph i1.pcx}
\vspace{20cm}
\end{figure}

\end{document}